\documentstyle[12pt]{article}
\begin{document}
\title{The Cosmology of Fluctuations}
\author{B.G.Sidharth \\
Centr for Applicable Mathematics and Computer Sciences\\
B.M.Birla Science Centre,Hyderabad, 500463,India}
\date{}
\maketitle
\begin{abstract}
We review a cosmology in which particles are fluctuationally created from
a background Zero Point Field. This cosmology is consistent with recent
observations of an ever expanding and accelerating universe, as also the
recently confirmed evolution of the fine structure constant. All hitherto
mysterious and accidental, so called Large Number coincidences, infact follow
from the theory.
\end{abstract}
\section{Introduction}
From early 1998, the conventional wisdom of cosmology that had concretized from
the mid sixties onwards, began to be challenged. It had been believed that the
density of the universe is near its critical value, separating eternal expansion
and ultimate contraction, while the nuances of the dark matter theories were
being fine tuned. However the work of Perlmutter and others \cite{r1,r2}
began appearing in 1998 and told a different story. These observations
of distant supernovae indicated that contrary to widely held belief, the universe
was not only not decelarating, it was actually accelerating. This paradigm shift
permeated to the popular press also. For example an article in the Scientific
American \cite{r3} observed, "In recent years the field
of cosmology has gone through a radical upheaval. New discoveries have
challenged long held theories about the evolution of the universe... Now that
observers have made a strong case for cosmic acceleration, theorists must
explain it.... If the recent turmoil is anything to go by, we had better keep
our options open."\\
On the other hand, the Physics World observed \cite{r4},
"A revolution is taking place in cosmology. New ideas are usurping traditional
notions about the composition of the universe, the relationship between geometry
and destiny, and Einstein's greatest blunder."\\
The infamous cosmological constant was resurrected and now it was "dark energy"
that was in the air, rather than dark matter.\\
Shortly before these dramatic discoveries, the author had presented a cosmological
model based on fluctuations in an all permeating Zero Point Field
\cite{r5,r6,r7,r8}. This model is consistent with astrophysical observations
and predicts an ever expanding and accelerating universe. It deduces from
theory the so called large number coincidences including the purely empirical
Weinberg formula that connects the pion mass to the Hubble Constant \cite{r9,r10}.
Let us now examine this cosmology and some of its implications.
\section{Fluctuations and Cosmology}
We first observe that the concept of a Zero Point Field (ZPF) or Quantum Vacuum
(or Ether) is an idea whose origin can be traced back to Max Planck himself. Quantum
Field Theory attributes the ZPF to the virtual Quantum Effects of an already
present electromagnetic field \cite{r11}. There is another approach,
sometimes called Stochastic Electrodynamics which treats the ZPF as primary
and attributes to it Quantum Mechanical effects \cite{r12,r13}. It may be
observed that the ZPF results in the well known experimentally verified
Casimir effect \cite{r14,r15}. We would also like to point out that
contrary to popular belief, the concept of Ether has survived over the decades
through the works of Dirac, Vigier, Prigogine, String Theoriests like Wilzeck
and others \cite{r16}-\cite{r24}. It appears that even Einstein himself
continued to believe in this concept \cite{r25}.\\
We would first like to observe
that the energy of the fluctuations in the background electromagnetic field could lead
to the formation of elementary particles. Infact it is known that this energy
of fluctuation in a region of length $l$ is given by \cite{r26}
$$B^2 \sim \frac{\hbar c}{l^4}$$
In the above if $l$ is taken to be the Compton wavelength of a typical elementary
particle, then we recover its energy $mc^2$, as can be easily verified. It may
be mentioned that Einstein himself had believed that the electron was a result
of such condensation from the background electromagnetic field (Cf.\cite{r27,r7,r8}
for details). We also take the pion to represent a typical elementary
particle, as in the literature.\\
To proceed, as there are $N \sim 10^{80}$ such particles in the universe,
we get
\begin{equation}
Nm = M\label{e1}
\end{equation}
where $M$ is the mass of the universe.\\
In the following we will use $N$ as the sole cosmological parameter.\\
Equating the gravitational potential energy of the pion in a three dimensional isotropic
sphere of pions of radius $R$, the radius of the universe, with the rest
energy of the pion, we can deduce the well known relation \cite{r28,r29}
\begin{equation}
R \approx \frac{GM}{c^2}\label{e2}
\end{equation}
where $M$ can be obtained from (\ref{e1}).\\
We now use the fact that given $N$ particles, the fluctuation in the particle number is of the
order $\sqrt{N}$\cite{r29,r30,r7,r8,r5,r6}, while a typical time interval for the
fluctuations is $\sim \hbar/mc^2$, the Compton time. We will come back to
this point later. So we have
$$\frac{dN}{dt} = \frac{\sqrt{N}}{\tau}$$
whence on integration we get,
\begin{equation}
T = \frac{\hbar}{mc^2} \sqrt{N}\label{e3}
\end{equation}
We can easily verify that equation (\ref{e3}) is indeed satisfied
where $T$ is the age of the universe. Next by differentiating (\ref{e2}) with
respect to $t$ we get
\begin{equation}
\frac{dR}{dt} \approx HR\label{e4}
\end{equation}
where $H$ in (\ref{e4}) can be identified with the Hubble Constant, and using
(\ref{e2}) is given by,
\begin{equation}
H = \frac{Gm^3c}{\hbar^2}\label{e5}
\end{equation}
Equation (\ref{e1}), (\ref{e2}) and (\ref{e3}) show that in this formulation, the correct mass,
radius and age of the universe can be deduced given $N$ as the sole
cosmological or large scale parameter. Equation (\ref{e5}) can be written as
\begin{equation}
m \approx \left(\frac{H\hbar^2}{Gc}\right)^{\frac{1}{3}}\label{e6}
\end{equation}
Equation (\ref{e6}) has been empirically known as an "accidental" or "mysterious" relation.
As observed by Weinberg\cite{r10}, this is unexplained: it relates a single
cosmological parameter $H$ to constants from microphysics. We will touch upon
this micro-macro nexus again.
In our formulation, equation (\ref{e6}) is no longer a mysterious coincidence but
rather a consequence.\\
As (\ref{e5}) and (\ref{e4}) are not exact equations but rather,
order of magnitude relations, it follows that a
small cosmological constant $\wedge$ is allowed such that
$$\wedge \leq 0 (H^2)$$
This is consistent with observation and shows that $\wedge$ is very very small - this has been a puzzle,
the so called cosmological constant problem \cite{r31}. But it is explained
here.\\
To proceed we observe that because of the fluctuation of $\sim \sqrt{N}$ (due to the ZPF),
there is an excess electrical potential energy of the electron, which infact
we have identified as its inertial energy. That is \cite{r7,r29},
$$\sqrt{N} e^2/R \approx mc^2.$$
On using (\ref{e2}) in the above, we recover the well known Gravitation-electromagnetism
ratio viz.,
\begin{equation}
e^2/Gm^2 \sim \sqrt{N} \approx 10^{40}\label{e7}
\end{equation}
or without using (\ref{e2}), we get, instead, the well known so called
Eddington formula,
\begin{equation}
R = \sqrt{N}l\label{e8}
\end{equation}
Infact (\ref{e8}) is the spatial counterpart of (\ref{e3}). If we combine (\ref{e8}) and (\ref{e2}), we get,
\begin{equation}
\frac{Gm}{lc^2} = \frac{1}{\sqrt{N}} \propto T^{-1}\label{e9}
\end{equation}
where in (\ref{e9}), we have used (\ref{e3}). Following Dirac (cf.also \cite{r32})
we treat $G$ as the variable, rather than the quantities $m, l, c \mbox{and}
\hbar$ (which we will call micro physical constants) because of their central role
in atomic (and sub atomic) physics.\\
Next if we use $G$ from (\ref{e9}) in (\ref{e5}), we can see that
\begin{equation}
H = \frac{c}{l} \quad \frac{1}{\sqrt{N}}\label{e10}
\end{equation}
Thus apart from the fact that $H$ has the same inverse time dependance on
$T$ as $G$, (\ref{e10}) shows that given the microphysical constants, and
$N$, we can deduce the Hubble Constant also, as from (\ref{e10}) or (\ref{e5}).\\
Using (\ref{e1}) and (\ref{e2}), we can now deduce that
\begin{equation}
\rho \approx \frac{m}{l^3} \quad \frac{1}{\sqrt{N}}\label{e11}
\end{equation}
Next (\ref{e8}) and (\ref{e3}) give,
\begin{equation}
R = cT\label{e12}
\end{equation}
(\ref{e11}) and (\ref{e12}) are consistent with observation.\\
Finally, we observe that using $M,G \mbox{and} H$ from the above, we get
\begin{equation}
M = \frac{c^3}{GH}\label{e13}
\end{equation}
The relation (\ref{e13}) is required in the Friedman model of the expanding universe
(and the Steady State model also).\\
The above model predicts an ever expanding and possibly accelerating universe
whose density keeps decreasing. This seemed to go against the
accepted idea that the density of the universe equalled the critical density
required for closure.
\section{Issues and Ramifications}
i) The above cosmology exhibits a time variation of the gravitational constant of the
form
\begin{equation}
G = \frac{\beta}{T}\label{e14}
\end{equation}
Indeed this is true in a few other schemes also, including Dirac's cosmology
(Cf. \cite{r33,r34,r27}). Interestingly it can be shown that such a
time variation can explain the precession of the perihelion of Mercury (Cf.
\cite{r35}). It can also provide an alternative explanation for dark
matter and the bending of light while the Cosmic Microwave Background
Radiation is also explained (Cf.\cite{r27}).\\
It is also possible to deduce the existence of gravitational waves given
(\ref{e14}). To see this quickly let us consider the Poisson equation for the
metric $g_{\mu \nu}$
\begin{equation}
\nabla^2 g_{\mu \nu} = G \rho u_\mu u_\nu\label{e15}
\end{equation}
The solution of (\ref{e15}) is given by
\begin{equation}
g_{\mu \nu} = G \int \frac{\rho u_\mu u_\nu}{|\vec r - \vec r'|}
d^3\vec r\label{e16}
\end{equation}
Indeed equations similar to (\ref{e15}) and (\ref{e16}) hold for the
Newtonian gravitational potential also. If we use the second time derivative of
$G$ from (\ref{e14}) in (\ref{e16}), along with (\ref{e15}), we can
immediately obtain the D'alembertian wave equation for gravitational waves,
instead of the Poisson equation:
$$\Box g_{\mu \nu} \approx 0$$
ii) Recently a small variation with time of the fine structure constant has
been detected and reconfirmed by Webb and coworkers \cite{r36,r37}. This observation is
consistent with the above cosmology. We can see this as follows. We use an
equation due to Kuhne \cite{r38}
\begin{equation}
\frac{\dot \alpha_z}{\alpha_z} = \alpha_z \frac{\dot H_z}{H_z},\label{e17}
\end{equation}
If we now use the fact that the cosmological constant $\Lambda$ is given by
\begin{equation}
\Lambda \leq 0(H^2)\label{e18}
\end{equation}
as can be seen from (\ref{e4}), in (\ref{e17}), we get using (\ref{e18}),
\begin{equation}
\frac{\dot \alpha_z}{\alpha_z} = \beta H_z\label{e19}
\end{equation}
where $\beta < - \alpha_z < - 10^{-2}$.\\
Equation (\ref{e19}) can be shown to be the same as
\begin{equation}
\frac{\dot \alpha_z}{\alpha_z} \approx -1 \times 10^{-5} H_z.\label{e20}
\end{equation}
which is the same as Webb's result.\\
We give another derivation of (\ref{e20}) in the above context wherein, as the
number of particles in the universe increases with time, we go from the Planck
scale to the Compton scale.\\
This can be seen as follows: In equation (\ref{e7}), if the number of particles
in the universe, $N = 1$, then the mass $m$ would be the Planck mass. In this
case the classical Schwarzschild radius of the Planck mass would equal its
Quantum Mechanical Compton wavelength. To put it another way, all the energy
would be gravitational (Cf.\cite{r27} for details). However as the number of particles
$N$ increases with time, according to (\ref{e3}), gravitation and electromagnetism
get differentiated and we get (\ref{e7}) and the Compton scale.\\
It is known that the Compton length, due to
zitterbewegung causes a correction to the electrostatic potential which an
orbiting electron experiences, rather like the Darwin term \cite{r11}.\\
Infact we have
$$\langle \delta V \rangle = \langle V (\vec r + \delta \vec r)\rangle - V
\langle (\vec r )\rangle$$
$$= \langle \delta r \frac{\partial V}{\partial r} + \frac{1}{2} \sum_{\imath j}
\delta r_\imath \delta r_j \frac{\partial^2 V}{\partial r_\imath \partial r_j} \rangle$$
\begin{equation}
\approx 0(1) \delta r^2 \nabla^2 V\label{e21}
\end{equation}
Remembering that $V = e^2/r$ where $r \sim 10^{-8}cm$,
from (\ref{e21}) it follows that if $\delta r \sim l$, the Compton wavelength
then
\begin{equation}
\frac{\Delta \alpha}{\alpha} \sim 10^{-5}\label{e22}
\end{equation}
where $\Delta \alpha$ is the change
in the fine structure constant from the early universe. (\ref{e22}) is an
equivalent form of (\ref{e20}) (Cf.ref.\cite{r38}), and is the result originally
obtained by Webb et al (Cf.refs.\cite{r36,r37}).\\
iii) The latest observations of distant supernovae referred to above indicate
that the closure parameter $\Omega \leq 1$.\\
Remembering that $\Omega$ is given by \cite{r39}
$$\Omega = \frac{8\pi G}{3 H^2} \rho$$
we get therefrom on using (\ref{e1})
$$\frac{H^2}{2G} R^3 = mN$$
which immediately leads to the mysterious Weinberg formula (\ref{e6}). Thus
this is the balance between the cosmos at large and the micro cosmos.\\
iv) In General Relativity as well as in the Newtonian Theory, we have,
without a cosmological constant
\begin{equation}
\ddot{R} = -\frac{4}{3} \pi G \rho R\label{e23}
\end{equation}
We remember that there is an uncertainity in time to the extent of the Compton
time $\tau$, and also if we now use the fact that $G$ varies with time,
(\ref{e23}) becomes on using (\ref{e14}),
$$\ddot{R} = -\frac{4}{3} \pi G (t - \tau) \rho R$$
\begin{equation}
= -\frac{4}{3} \pi G \rho R + \frac{4}{3} \pi \rho R \left(\frac{\tau}{t}\right)
G/t\label{e24}
\end{equation}
Remembering that at any point of time, the age of the universe, that is
$t$ itself is given by (\ref{e3}), we can see from (\ref{e24}) that this
effect of time variation of $G$, which again is due to the background Zero
Point Field is the same as an additional density, the vacuum density given
by
\begin{equation}
\rho_{vac} = \frac{\rho}{\sqrt{N}}\label{e25}
\end{equation}
This term in (\ref{e24}) is also equivalent to the presence of a cosmological
constant $\Lambda$ as discussed above. On the other hand, we know independently
that the presence of a vacuum field leads to a cosmological constant given
by (Cf.ref.\cite{r27} and references therein)
\begin{equation}
\Lambda = G \rho_{vac}\label{e26}
\end{equation}
Equation (\ref{e26}) is pleasingly in agreement with (\ref{e24}) and (\ref{e25}) that is, the
preceeding considerations. In other words quantitatively we have reconfirmed
that it is the background Zero Point Field that manifests itself as the
cosmological constant described in Section 2. This also gives as pointed out an explanation
for the so called cosmological constant problem \cite{r31} viz., why is
the cosmological constant so small?\\
v) In the above cosmology of fluctuations, our starting point was the creation
of $\sqrt{N}$ particles within the minimum time interval, a typical elementary
particle Compton time $\tau$. A rationale for this, very much in the spirit of
the condensation of particles from a background Zero Point Field as discussed
at the beginning of Section 2, has also been obtained recently in terms of a
broken symmetry phase transition from the Zero Point Field or Quantum Vacuum.
In this case, particles are like the Benard cells which form in fluids, as a result
of a phase transition. While some of the particles or cells may revert to
the Zero Point Field, on the whole there is a creation of these particles. If
the average time for the creation of one of these particles or cells is
$\tau$, then at any point of time where there are $N$ such particles, the time
elapsed, in our case the age of the universe, would be given by (\ref{e3}) (Cf.
\cite{r40}). While this is not exactly the Big Bang scenario, there is
nevertheless a rapid creation of matter from the background Quantum Vacuum
or Zero Point Field. Thus half the matter of the universe would have been
created within a fraction of a second.\\
In any case when $\tau \to 0$, we recover the Big Bang scenario with a singular
creation of matter, while when $\tau \to$ Planck time we recover the Prigogine
Cosmology (Cf.\cite{r27} for details). However in neither of these two limits we can
deduce all the above consistent with observation relations.\\
vi) The above cosmological model is related to the fact that there are minimum
space time intervals $l, \tau$. Indeed in this case it is known that there is
an underlying non commutative geometry of spacetime \cite{r41,r42,r43} given by
\begin{equation}
[x,y] \approx 0(l^2),[x,p_x] = \imath \hbar [1 + \beta l^2], [t,E] = \imath \hbar
[1+ \gamma \tau^2]\label{e27}
\end{equation}
Interestingly (\ref{e27}) implies modification to the usual Uncertainity Principle,
and this in turn  can also be interpreted in terms of a variable speed of light
cosmology \cite{r44,r45,r46}.\\
The relations (\ref{e27}), lead to the modified Uncertainity relation
\begin{equation}
\Delta x \sim \frac{\hbar}{\Delta p} + \alpha' \frac{\Delta p}{\hbar}\label{e28}
\end{equation}
(\ref{e28}) appears also in Quantum SuperString Theory and is related to the
well known Duality relation
$$R \to \alpha'/ R$$
(Cf.\cite{r47,r48}). In any case (\ref{e28}) is symptomatic of the fact
that we cannot go down to arbitrarily small space time intervals. We observe
that the first term of (\ref{e28}) gives the usual Uncertainity relation. In
the second term,
we write $\Delta p = \Delta Nmc$,
where $\Delta N$ is the Uncertainity in the number of particles, $N$, in the
universe. Also $\Delta x = R$, the radius of the universe where
$$
R \sim \sqrt{N}l,$$
the famous Eddington relationship. It should be stressed that the otherwise
emperical Eddington formula, arises quite naturally in a Brownian characterisation
of the universe as has been pointed out earlier (Cf. for example ref.\cite{r49}). Put
simply (\ref{e8}) is the Random Walk equation\\
We now get,
$$\Delta N = \sqrt{N}$$
This is the uncertainity in the particle number, we used earlier.
Substituting this in the time analogue of the second term of (\ref{e28}), we
immediately get, $T$ being the age of the universe,
$$T = \sqrt{N} \tau$$
which is equation (\ref{e3}). So, our cosmology is self consistent with the
modified relation (\ref{e28}).\\
Interestingly these minimum space time considerations can be related to the
Feynmann-Wheeler Instantaneous Action At a Distance formulation (Cf.\cite{r50,r51,r52}).\\
We finally remark that relations like (\ref{e27}) and (\ref{e28}), which can
also be expressed in the form, $a$ being the minimum length,
$$[x, p_x] = \imath \hbar [1 + \left(\frac{a}{\hbar}\right)^2 p^2]$$
(and can be considered to be truncated from a full series on the right hand side
(Cf. \cite{r53}), could be deduced from the rather simple model of a fixed
lattice - a one dimensional lattice for simplicity. In this case we will have
(Cf.\cite{r27})
$$[x,p_x] = \imath \hbar cos \left(\frac{p}{\hbar} a\right),$$
where $a$ is the lattice length, $l$ the Compton length in our case. The energy
time relation now leads to a correction to the mass energy formula, viz
$$E = mc^2 cos (kl), k \equiv p/\hbar$$
This is the contribution of the extra term in the Uncertainity Principle.\\
vii) It is well known that the Planck scale is an absolute minimum scale in
the universe. In Section 3, ii) we argued that with the passage of time the
Planck scale would evolve to the present day elementary particle Compton
scale. This can also be seen in the following way: We have by definition
$$\hbar G/c^3 = l^2_P$$
where $l_P$ is the Planck length $\sim 10^{-33}cms$. If we use (\ref{e9})
in the above we will get
\begin{equation}
l = N^{1/4} l_P\label{e29}
\end{equation}
Similarly we have
\begin{equation}
\tau = N^{1/4} \tau_P\label{e30}
\end{equation}
In (\ref{e29}) and (\ref{e30}) $l$ and $\tau$ denote the typical elementary
particle Compton length and time scale, and $N$ is the number of such elementary
particles in the universe. We could explain these equations in terms of the
Benard cell like elementary particles referred to above. This time there are
total of $n = \sqrt{N}$ Planck particles and (\ref{e29}) and (\ref{e30}) are
the analogues of equations (\ref{e3}) and (\ref{e8}) in the context of the
formation of such particles. Indeed it is well known that a Planck mass,
$m_P \sim 10^{-5}gms$, has a Compton life time and also a Bekenstein Radiation
life time of the order of the Planck time. These space time scales are much
too small and we encounter much too large energies from the point of view of
our observed limits. As noted above our observed scale is the Compton scale,
in which Planck scale phenomena are moderated. In any case it can be seen from
the above that as the number of particles $N$ increases, the scale evolves
from the Planck to the Compton scale. Interestingly another way to looking at the above
is that the particles can be considered to be the fluctuational effect of
the fluctuationally created $\sqrt{N}$ particles (Cf.\cite{r27}).\\
So, the scenario which emerges is, that as the universe evolves, Planck particles
form the underpinning for elementary particles, which in turn form the
underpinning for the universe by being formed continuously.\\
This can be confirmed by the following argument: We can rewrite (\ref{e29})
as
\begin{equation}
l = \nu' \sqrt{T}\label{e31}
\end{equation}
$$\nu' = l_P/\sqrt{\tau} \approx \hbar/m_P$$
wherein we have used (\ref{e3}). Equation (\ref{e31}) is identical to the
Nelsonian-Brownian Theory which is infact the underpinning for equations like
(\ref{e3}) or (\ref{e8}), except that this time we have the same Brownian
Theory operating from the Planck scale to the Compton scale, instead of
from the Compton scale to the edge of the universe as seen above (Cf. also
\cite{r49,r27}).\\
Interestingly, let us apply the above scenario of $\sqrt{n}$ Planck particles
forming an elementary particle, to the extra term of the modified Uncertainity
Principle (\ref{e28}), as we did earlier in section (iv). Remembering that
$\alpha' = l^2_P$ in the theory, and $\Delta p = N^{1/4} m_P c$, in this case,
we get, as $\Delta x = l$,
$$l = N^{1/4} l_P,$$
which will be recognized as (\ref{e29}) itself! Thus once again we see how the
above cosmology is consistently tied up with the non commutative space time
expressed by equations (\ref{e27}) or (\ref{e28}).\\
It may be mentioned that, as indeed can be seen from (\ref{e29}) and (\ref{e30}),
in this model, the velocity of light remains constant.


\begin{thebibliography}{99}
\bibitem {r1} S. Perlmutter, et al, Nature, Vol.391, 1 January 1998, p.51-59.
\bibitem {r2} R.P. Kirshner, Proc. Natl. Acad. Sci. USA, Vol.96, April 1999,
pp.4224-4227.
\bibitem {r3} G. Musser and M. Alpert, Scientific American, January 2000, pg.27.
\bibitem {r4} R.R. Caldwell and P.J. Steinhardt, Physics World, November 2000, pg.31.
\bibitem {r5} B.G. Sidharth, in "Frontiers of Quantum Physics",
Eds., Lim, S.C., et al, Springer Verlag, Singapore, 1998.
\bibitem {r6} B.G. Sidharth, Proc. of the Eighth Marcell Grossmann Meeting on
General Relativity, Ed. T. Piran, World Scientific, Singapore, 1999, p.476-479.
\bibitem {r7} B.G. Sidharth, Int.J. of Mod.Phys.A 13(15), 1998, pp2599ff.
\bibitem {r8} B.G. Sidharth, International Journal of Theoretical
Physics, Vol.37, No.4, 1998, 1307-1312.
\bibitem {r9} J.V. Narlikar, "Introduction to Cosmology", Cambridge University
Press, Cambridge, 1993, pp.237ff.
\bibitem {r10} S. Weinberg, "Gravitation and Cosmology", John Wiley \& Sons,
New York, 1972, p.62.
\bibitem {r11} J.D. Bjorken, and S.D. Drell, "Relativistic Quantum Mechanics",
Mc-Graw Hill, New York, 1964, p.39.
\bibitem {r12} E. Santos, "Stochastic Electrodynamics and the Bell Inequalities"
in "Open Questions in Quantum Physics", Ed. G. Tarozzi and A. van der Merwe,
D. Reidel Publishing Company, 1985, p.283-296.
\bibitem {r13} L. De Pena, "Stochastic Processes applied to Physics...", Ed.,
B Gomez, World Scientific, Singapore, 1983.
\bibitem {r14} V.M. Mostepanenko and N.N. Trunov, Sov.Phys.Usp. 31(11), November
1988, p.965-987.
\bibitem {r15} S.K. Lamoreauz, Phys.Rev.Lett., Vol.78, No.1, January 1997, p.5-8.
\bibitem {r16} N.C. Petroni and J.P. Vigier, Foundations of Physics, Vol.13, No.2,
1983, p.253-286.
\bibitem {r17} M.T. Raiford, Physics Today, July 1999, p.81.
\bibitem {r18} P.W. Milonni, Physica Scripta. Vol. T21, 1988, p.102-109.
\bibitem {r19} P.W. Milonni and M.L. Shih, Am.J.Phys., 59 (8), 1991, p.684-698.
\bibitem {r20} F. Wilczek, Physics Today, November 1999, p.11-13.
\bibitem {r21} F. Wilczek, Physics Today, January 1999, p.11-13.
\bibitem {r22} T.D. Lee, "Statistical Mechanics of Quarks and Hadrons", Ed. H. Satz,
North-Holland Publishing Company, 1981, p.3ff.
\bibitem {r23} V. Hushwater, Am.J.Phys. 65(5), May 1997, p.381--384.
\bibitem {r24} P.W. Milonni, "The Quantum Vacuum An Introduction to
Quantum Electrodynamics", Academic Press, New York, 1994.
\bibitem {r25} P. Achuthan et al, in "Gravitation, Quanta and the Universe",
Ed. A.R. Prasanna, J.V. Narlikar, C.V. Vishveshwara, Wiley Eastern, New Delhi,
1980, p.300.
\bibitem {r26} C.W. Misner, K.S. Thorne and J.A. Wheeler, "Gravitation",
W.H. Freeman, San Francisco, 1973, p.1190ff.
\bibitem {r27} B.G. Sidharth, "Chaotic Universe: From the Planck to the Hubble Scale",
Nova Science Publishers, New York, 2001, p.20.
\bibitem {r28} L. Nottale, "Fractal Space-Time and Microphysics: Towards
a Theory of Scale Relativity", World Scientific, Singapore, 1993, p.312.
\bibitem {r29} S. Hayakawa,  Suppl of PTP Commemmorative Issue, 1965, 532-541.
\bibitem {r30} K. Huang, "Statistical Mechanics", Wiley Eastern, New Delhi 1975, pp.75ff.
\bibitem {r31} S. Weinberg, Phys.Rev.Lett, 43, 1979, p.1566.
\bibitem {r32} V.N. Melnikov, International Journal of Theoretical Physics,
\underline{33} (7), 1994, 1569-1579.
\bibitem {r33} J.V. Narlikar, Foundations of Physics, Vol.13, No.3, 1983, p.311-323.
\bibitem {r34} J.D. Barrow and P. Parsons, Phys.Rev.D., Vol.55, No.4, 15 February 1997,
p.1906ff.
\bibitem {r35} B.G. Sidharth, Il Nuovo Cimento, 115B (12), (2), 2000, pg.151.
\bibitem {r36} J.K. Webb, et al., Phys.Rev.Lett., 87 (9), 2001, pp.091301-1 ff.
\bibitem {r37} J.K. Webb, et al., Phys.Rev.Lett., 82, 1999, pp884ff.
\bibitem {r38} R.W. Kuhne, Mod.Phys.Lett.A, Vol.14, No.27, 1999, pp.1917-1922.
\bibitem {r39} C.H. Ohanian, and R. Ruffini, "Gravitation and Spacetime",
New York, 1994, pp.130ff.
\bibitem {r40} B.G. Sidharth, "The Nature of Quantum Space Time", to appear in
Chaos, Solitons and Fractals.
\bibitem {r41} B.G. Sidharth, Chaos, Solitons and Fractals, 13, 2002, p.1369-1370.
\bibitem {r42} B.G. Sidharth, Il Nuovo Cimento, 116B (6), 2001, pg.4 ff.
\bibitem {r43} B.G. Sidharth, Frontiers of Fundamental Physics 4, Plenum
Publishers/Kluwer Academic, New York, 2001, p.97-107.
\bibitem {r44} S. Kalyana Rama, Phys.Lett. B, 519, 2001, p.103-110.
\bibitem {r45} J.W. Moffat, Int.J.Mod.Phys.,D 2, 1993, p.351, J.W. Moffat,
Found.Phys. 23, 1993.
\bibitem {r46} J. Magueijo, Phys.Rev.D., 62, 2000, 103521.
\bibitem {r47} W. Witten, Physics Today, April 1996, pp.24-30.
\bibitem {r48} G. Veneziano in "The Geometric Universe", Ed. by
S.A. Huggett, et al.,  Oxford University Press, Oxford, 1998, p.235ff.
\bibitem {r49} B.G. Sidharth, Chaos, Solitons and Fractals, 12, 2001, p.173-178.
\bibitem {r50} F. Hoyle and J.V. Narlikar, The Astrophysical Journal, 410, June
20 1993, p.437-457.
\bibitem {r51} B.G. Sidharth, in Instantaneous Action at a Distance in
Modern Physics: "Pro and Contra" , Eds., A.E. Chubykalo et. al., Nova Science
Publishing, New York, 1999.
\bibitem {r52} B.G Sidharth, "A Reconciliation of Electromagnetism and
Gravitation", to appear in  Annales de Fondation De Broglie, in press.
\bibitem {r53} Y. Ne'eman, in Proceedings of the First Internatioinal Symposium,
"Frontiers of Fundmental Physics", Eds. B.G. Sidharth and A. Burinskii,
Universities Press, Hyderabad, 1999, pp.83ff.
\end{thebibliography}
\end{document}